\documentstyle[epsf,11pt,aaspp4]{article}

\begin{document}
\title{Temporal Properties of Cygnus X--1 During the Spectral Transitions}
\author{Wei~Cui\altaffilmark{1}, S.~N.~Zhang\altaffilmark{2}, W.~Focke\altaffilmark{3}\altaffilmark{4}, and J.~H.~Swank\altaffilmark{3}}
\altaffiltext{1}{Room 37-571, Center for Space Research, Massachusetts Institute of Technology, Cambridge, MA 02139}
\altaffiltext{2}{ES-84, NASA/Marshall Space Flight Center, Huntsville, AL 35812}
\altaffiltext{3}{NASA/Goddard Space Flight Center, Code 662, Greenbelt, MD 20771}
\altaffiltext{4}{also Department of Physics, University of Maryland, College Park, MD 20741}
\authoremail{cui@space.mit.edu}
\slugcomment{Revised on 2/5/97}
\begin{abstract}

We report the results from our timing analysis of 15 {\it Rossi X--ray Timing
Explorer} (RXTE) observations of Cygnus~X--1 throughout its 1996 spectral 
transitions. The entire period can be divided into 3 distinct phases: (1) 
transition from the hard state to the soft state, (2) soft state, and (3) 
transition from the soft state back to the hard state. The observed X--ray
properties (both temporal and spectral) in Phases 1 and 3 are remarkably 
similar, suggesting that the same physical processes are likely involved in 
triggering such transitions. The power density spectrum (PDS) during the 
transition can be characterized by a low--frequency red noise (power law) 
component, followed by a white noise (flat) component which extends to 
roughly 1--3 Hz where it is cut off, and a steeper power law ($\sim 1/f^2$) 
at higher frequencies. The X--ray flux also exhibits apparent quasi--periodic 
oscillation (QPO) with the centroid frequency varying in the range of 4--12 
Hz. The QPO shows no correlation with the source flux, but becomes more
prominent at higher energies. This type of PDS bears resemblance to that of 
other black hole candidates often observed in a so--called very high state, 
although the origin of the observed QPO may be very different. The 
low--frequency red noise has not been observed in the hard state, thus 
seems to be positively correlated with the disk mass accretion rate which is 
presumably low in the hard state and high in the soft state; in fact, it 
completely dominates the PDS in the soft state. In the framework of thermal
Comptonization models, Cui et al. (1997a) speculated that the difference in 
the observed spectral and timing properties between the hard and soft states 
is due to the presence of a ``fluctuating'' Comptonizing corona during the 
transition. Here we present the measured hard X--ray time lags and coherence 
functions between various energy bands, and show that the results strongly 
support such scenario.

\end{abstract}

\keywords{binaries: general --- stars: individual (Cygnus~X--1) --- X--rays: stars}

\section{INTRODUCTION}

Cygnus~X--1 is one of the best studied X--ray sources. It was discovered
in 1965 (\cite{bowyer1965}), and its binary nature was soon established 
with the detection of an orbital period of 5.6 days (\cite{bolton1972}; 
\cite{webster1972}). The optical radial--velocity measurements indicate 
that the compact object has a mass in excess of $\sim 7 M_{\odot}$ and a 
probable mass of $\sim 16 M_{\odot}$ (Gies \& Bolton 1982, 1986),
strongly suggesting that there is a black hole in the system.

Cyg~X--1 belongs to the class of high--mass X--ray binaries. Its companion was
identified as a O9.7 Iab supergiant with a mass in excess of $\sim 20 
M_{\odot}$ and a probable mass of $\sim 33 M_{\odot}$ (\cite{gies1986}).
Stellar wind is, therefore, postulated to play an important role in producing
X--rays. It is thought that the companion star is very close to filling its
Roche Robe, and the X--ray emission is driven by so--called ``focused wind 
accretion'' (\cite{gies1986}).

The long--term monitoring of Cyg~X--1 revealed that its soft X--ray flux 
($\lesssim$ 10 keV) shows, on average, two distinct levels (\cite{holt1976};
\cite{cuietal1997a}), which are often referred to as the low and high states 
in the literature for historical reasons. Such terminology can often be very
confusing because the hard X--ray flux ($\gtrsim$ 10 keV) is anti--correlated 
with the soft flux during the transition --- the soft X--ray low state is 
actually hard X--ray high (\cite{dolan1977}; \cite{ling1987}; 
\cite{cuietal1997a}). 
The observed X--ray properties of Cyg~X--1 depend strongly on which state it 
is in. It is usually in the soft X--ray low state where the energy spectrum is 
relatively flat (or hard), and can be characterized by a single power law 
with a photon index of $\sim 1.7$ (cf. review by \cite{tanaka1995}). 
Occasionally, it reaches the soft X--ray high state where the power--law energy 
spectrum becomes significantly steeper (or softer; with a photon index of 
$\sim 2.5$). Therefore, ``hard state'' and ``soft state'' are more precise
terms in describing the bi--modal behavior. 
Despite extensive investigation (observational and theoretical) for the past 
couple of decades, it is still not clear what triggers the spectral 
transitions. 

The X--ray spectrum of Cyg~X--1 extends beyond 100 keV in both states (e.g., 
\cite{phlipsetal1996}; \cite{cuietal1997a}), which presents challenge to 
theoretical models. It is generally thought that the hard X--ray emission is 
the result of low--energy photons being up--scattered by hot electrons in the 
system (e.g., \cite{shapiro1976}; review by \cite{narayan1996b}, and 
references therein). The soft photons probably originate in the synchrotron 
emission from relativistic electrons or the thermal 
emission from an accretion disk. However, little is known about the origin of 
Comptonizing electrons. Thermal Comptonization requires extremely high 
electron temperature ($\sim 10^{9}$ K) to reach the observed cutoff energy
(\cite{sunyaev1980}; \cite{payne1980}). The preliminary results from a recent 
simultaneous RXTE/OSSE observation of Cyg~X--1 in the soft state show that
the power--law energy spectrum extends up to $\sim$ 600 keV without any breaks 
(\cite{phlipsetal1997}). If confirmed, they would be at odds with those that 
require a ``cooled'' Comptonizing region in the soft state, and push others 
to the physical limits as well. Alternatively, in the soft state, the bulk 
motion of relativistic electrons near the black hole may be more efficient 
in up--scattering soft photons than the thermal motion (\cite{titarchuketal1996}). 
Calculations show that such bulk motion can produce the observed spectral 
characteristics (\cite{chakrabarti1995}; \cite{ebisawa1996}; 
\cite{titarchuk1997}). 

There is, however, a tendency that the temporal properties are not emphasized 
enough and sometimes are completely ignored in the model. X--ray variability 
carries rich information about X--ray emitting regions, instabilities in the 
accretion disk, temporal evolution of the disk and Comptonizing region during 
the transition, and even the condition about the companion star (e.g., Roche 
Robe overflow versus focused wind accretion). Of course, high signal--to--noise 
and high timing resolution data are essential for such studies, especially 
phenomena in the vicinity of the black hole where dynamical time scales are 
on the order of tens or hundreds of microseconds. The RXTE data have much
better simultaneous spectral and timing information than was previously 
available.

RXTE provides unprecedented true $\mu s$ timing resolution and covers a broad 
energy range (\cite{bradt1993}). The {\it All--Sky Monitor} (ASM; 
\cite{levine1996}) aboard routinely monitors known bright sources (in the 
1.3--12 keV energy band). On 1996 May 10 (MJD 50213), it revealed that Cyg~X--1 
started a transition from the normal hard state to the soft state 
(\cite{cui1996}; \cite{cuietal1996}). After reaching the soft state, it stayed
for nearly 2 month before going back down to the hard state 
(\cite{zhang1996}). Figure~1 shows the ASM light curve that covers the entire
period. 

This period can be divided into 3 distinct phases: (1) hard--to--soft
transition, (2) soft state, and (3) soft--to--hard transition. The transitions
are characterized by a fast rise (or decay) in the ASM flux. The soft--to--hard
transition is nearly a mirror image of the hard--to--soft transition. Although 
the ASM flux of the source is, on average, $\sim$0.4 Crab in the hard state 
and $\sim$1.1 Crab in the soft state, it varies greatly on time scales of 
minutes to days in both states, with a significantly larger amplitude in the 
soft state. 

Snapshots of Cyg~X--1 were taken with the main pointing detectors on RXTE, 
namely, the {\it Proportional Counter Array} (PCA; 2--60 keV) and 
{\it High-energy X--ray Timing Experiment} (HEXTE; 15--250 keV), to monitor 
its temporal and spectral variability throughout the entire period. In this 
paper we present the results from the timing analysis of the PCA observations.
Preliminary spectral and timing results, based on the observations during the 
hard--to--soft transition and some in the soft state, were already reported 
in Belloni et al. (1996) and Cui et al. (1997a, b). For comparison, however, 
these observations were re--analyzed in 
the same way as subsequent ones for the soft state and the soft--to--hard 
transition. Table~1 briefly summarizes some basic information on the 
observations, which are also marked with symbols in Figure~1.

As discussed in Cui et al. (1997a,b) the source went through a sequence of 
states with the soft X-ray flux ($< 10$ keV) and the hard flux ($> 20$ 
keV) both changing. The soft flux alone was an incomplete indicator of 
the states. The RXTE observations with the pointing instruments were only 
snapshots. Systematic study of the spectral and timing properties during 
these snapshots and comparison of the BATSE and ASM behavior during the 
ASM soft outburst imply that the source went through a transition over a 
period of 20 days to a soft state which persisted for about 55 days and 
then made a transition back to the hard state again taking about 20 days. 

\section{ANALYSIS}

The PCA observations were made consistently with the following set of data 
modes (\cite{morgan1994}): a Binned mode with 4 ms time bin and 8 energy 
bands in the range of 2--13.1 keV, a Event mode with 16 $\mu s$ time bin and 
16 energy bands above 13.1 keV, and two Single--Bit modes with 122 $\mu s$ 
time bin, covering the energy bands 2--6.5 keV and 6.5--13.1 keV, respectively. 
The results presented here are derived from the Event--mode data and 
Single--Bit data. For brevity, the energy bands 2--6.5 keV, 6.5--13.1 keV, and 
13.1--60 keV are referred to as the soft band, medium band, and hard band, 
respectively.

Since all other observations contain only one orbit worth of data, we chose 
to break up the first and second observation (see Table~1) into individual 
orbits, each of which was analyzed separately. It should be noted that the 
first observation actually consists of only two orbits, but one of the five 
detectors was turned off (for safety reasons) during the second orbit. To 
simplify the analysis, the second orbit was broken up into two segments.

\subsection{Power Density Spectrum}

We chose to bin data in $2^{-6}$ s time bin. For each orbit, we broke up the 
light curve into 64--second segments. A $2^{12}$--point FFT was performed on 
each segment to obtain a Leahy--normalized PDS. The PDSs for all segments were 
then averaged to obtain the average PDS, as well as its variance. Finally, 
from the average PDS we subtracted the Poisson noise power corrected for 
instrument dead--time effects. The kind of effects of dead time caused by good
events were discussed by Zhang et al. (1995). Simulations show that for the 
real detectors, in which several effects compete (\cite{jahoda1996}), the 
effective dead-time can be described by the non-paralyzable formula (Equation 
(44) in Zhang et al. (1995)) with the detector dead-time $t_d \simeq 10\mu s$ 
(W. Zhang, private communication). In our case the time bin size $t_b=2^{-6}s$
and the formula is much simplified because $t_b >> t_d$ and the count rate is 
low (the time between events is large 
compared to the dead time). There is some spectral dependence to the dead time
and while we know the form and approximate numbers, we can expect that the 
numbers should be adjusted. A second effect needs to be taken into account (as
described in Appendix F of the XTE NRA and in Zhang et al. (1996)), the dead 
time due to very large events in the detector. No transmission of events is 
allowed for a set amount of time. For observations \#5-\#12 this was set to a 
value of about $70\mu s$, and for the others it was $155\mu s$. It is not a 
significant correction for the Cyg~X-1 observations. 

To estimate the uncertainty in calculating Poisson noise power, we obtained 
the PDSs that extend up to $\sim 4$ kHz by using finer time bins. We searched
but failed to detect any QPOs at frequencies above 30 Hz in all observations.
After carefully examining the PDS shape, it became apparent that the power in
the highest frequency bins was purely due to photon counting statistics. We 
then compared the measured high--frequency power to the calculated value, and 
concluded that roughly 99.5\% of the Poisson noise power should have been 
removed by following the dead time correction procedure we described. The 
residual becomes significant
starting at $\sim$30 Hz, which determined the finest time bin size we could 
use without losing information from PDS.

To facilitate the comparison of the results in different energy bands and 
from different observations, the power density is presented as the fractional
rms variability, which can be derived by dividing the Leahy--normalized PDS 
by the mean count rate (\cite{vandk1995}).

\subsubsection{Transitions}

The hard--to--soft transition was covered by the first 3 observations (a total 
of 7 orbits worth of data), while the soft--to--hard transition by the last 
three (see Table~1). In both phases, the observed PDS exhibits very similar 
characteristics. For the purpose of illustration, Figures~2 and 3 show the 
PDSs in the soft, medium, and hard band, as well as the total passing band, 
for Observations \#3 and \#15, respectively, representative of each phase. 
From the figures, the PDS can be characterized by a 
red noise component at low frequencies, followed by a white noise component 
that extends to 1--3 Hz, above which it is cut off. At higher frequencies, 
the PDS becomes power law again, with a much steeper slope (roughly -2, i.e., 
$1/f^2$). With respect to this model of continuum, a QPO is detected in all 
observations. To quantify the characteristics, we model the continuum with 
the following function:
\begin{equation}
PDS(f) = \left\{ 
\begin{array}{ll}
C(rms_c)^2 \left(\frac{f}{f_{b1}}\right)^{-\alpha_1} & \mbox{$f<f_{b1}$} \\
C(rms_c)^2 & \mbox{$f_{b1}\leq f < f_{b2}$} \\
C(rms_c)^2 \left(\frac{f}{f_{b2}}\right)^{-\alpha_2}& \mbox{$f\geq f_{b2}$},
\end{array}
\right.
\end{equation}
where C is a constant chosen so that $rms_c$ is the integrated fraction
rms amplitude of the continuum in the frequency range of 0.02--32 Hz. The 
QPO is modeled by a Lorentzian function, i.e.,
\begin{equation}
PDS(f) = \frac{(rms_{qpo})^2}{\pi} \frac{G/2}{(f-f_{qpo})^2 + (G/2)^2},
\end{equation}
where $rms_{qpo}$ is the integrated fractional rms amplitude of the QPO,
$f_{qpo}$ is the QPO centroid frequency, and G is the full width at half
maximum (FWHM). The best--fit models are summarized in Table~2.
The errors shown represent 1$\sigma$ confidence level. This simple model 
characterizes the data reasonably well, as indicated by the reduced $\chi^2$ 
of the fit.

The PDS continuum varies little on a time scale of days (the time between 
observations), except for the high--frequency cutoff ($f_{b2}$) which changes 
quite significantly. On a shorter time scale (between orbits), however, 
$f_{b2}$ remains quite stable. It is worth noting that the derived power--law 
slope of the red noise ($\alpha_1$) can be somewhat uncertain because of the 
narrow frequency range covered in the fit. It is actually closer to -1 (i.e., 
$1/f$ noise) when longer baselines are used (Cui et al. 1997a, b). Also shown 
in Figure~2 is that the continuum is energy dependent: the white noise power
decreases as the photon energy increases, while $f_{b2}$ remains roughly 
constant.

The QPO moves around significantly during the transition. There are no
apparent correlations between the QPO characteristics and the total count 
rate --- see Figure~4. However, the QPO is energy dependent: it becomes more 
prominent at higher energies, as shown in Figures~2 and 3. To investigate any
spectral dependence of the QPO, we define the following hardness ratios:
soft hardness ratio (HR1) is the ratio of the total source counts in the 
medium band to that in the soft band, and the hard hardness ratio (HR2)
the hard band to the medium band. Both hardness ratios were plotted against
the QPO fraction rms amplitude in Figure~5. With limited statistics, the QPO
appears to strengthen as the energy spectrum becomes harder.
Moreover, the QPO centroid frequency ($f_{qpo}$) shows strong 
anti--correlation with the fractional rms amplitude ($rms_{qpo}$) and 
correlation with $f_{b2}$ --- see Figure~6, which is made from Table~2.

\subsubsection{Soft State}

Starting from Observation \#4, subsequent 9 observations cover the soft 
state which is characterized by a power--law PDS (\cite{cuietal1996}; 
Cui et al. 1997a, b). As an example, Figure~7 shows the PDSs for Observation 
\#6 in the energy bands defined previously. The white noise is hardly 
detectable in this state, so the continuum is much simplified. It was 
modeled by a broken power law,
\begin{equation}
PDS(f) = \left\{ 
\begin{array}{ll}
C(rms_c)^2 \left(\frac{f}{f_b}\right)^{-\alpha_1} & \mbox{$f<f_b$} \\
C(rms_c)^2 \left(\frac{f}{f_b}\right)^{-\alpha_2}& \mbox{$f\geq f_b$}, 
\end{array}
\right.
\end{equation}
where C is a constant chosen so that $rms_c$ is the integrated fraction
rms amplitude of the continuum in the frequency range of 0.02--32 Hz. 
Examining Figure~7 more carefully, there appears to be some deviation from 
the power--law around 6 Hz. Such feature is so broad and can hardly be called
a QPO. For simplicity, these broad ``bumps'' are also modeled by the
Lorentzian function in Equation (2). The best--fit models are shown
in Table~3. In general, adding a QPO--like feature does statistically 
improve the fit, but such features are weak, and are not required at all in 
some cases. The centroid of these bumps clusters around 6--7 Hz. 

The continuum varies little in the soft state. The overall PDS is
dominated by the low--frequency 1/f noise, more so in the high
energy bands. The PDS breaks consistently at $\sim$ 13--14 Hz, except for 
Observations \#9 and \#12. At higher frequencies, the PDS steepens, with the
power--law slope ranging from -1.9 to -2.6, which is very similar to that 
during the transition. 

\subsection{Cross Spectral Correlation}

As in the previous section, we derived the average PDS and cross spectral
function (defined below) for each orbit worth of data by following the same 
procedure, except for a finer time bin size ($2^{-10}$ s) adopted and 
$2^{16}$--point FFT performed on each 64--second segment. It should be 
pointed out that the {\it unnormalized} PDS is used in the calculation. 

Because of the finer time bin size, much wider frequency range is covered. 
However, as noted in the previous section, the error in estimating Poisson 
noise power becomes significant above $\sim$32 Hz, so the results at high 
frequencies should be taken with reservation. 

Assuming that $F_1(f)$ and $F_2(f)$ are the Fourier series calculated for
two energy bands, \#1 and \#2, the cross spectral function (CSF) between 
them is defined as
\begin{equation}
\begin{array}{ll}
C(f) & = F^*_1(f) F_2(f),
\end{array}
\end{equation}
where $F^*_1(f)$ is the complex conjugate of $F_1(f)$. 

\subsubsection{Hard X--ray Time Lag}

If R(f) and I(f) are the real and imaginary parts of CSF, the average phase 
difference between the two energy bands is given by
\begin{equation}
\Delta \phi(f) = tan^{-1}\left(\frac{<I(f)>}{<R(f)>}\right),
\end{equation}
where angle brackets stand for ensemble averaging. Its variance was estimated 
by simply propagating errors, i.e., 
\begin{equation}
\delta \Delta \phi = 0.5\left|sin(2\Delta \phi)\right|\left(\left|\frac{\delta <R>}{<R>}\right|+\left|\frac{\delta <I>}{<I>}\right|\right).
\end{equation}
The average time lag (or advance if negative) of X--rays in energy band \#2
with respect to those in energy band \#1 is then given by
\begin{equation}
\Delta t = \frac{\Delta \phi}{2\pi f}.
\end{equation} 

The soft band is used as a reference band. For each of the observations 
listed in Table~1, the hard X--ray time lags are computed for the medium 
and hard bands with respect to the soft band. As for the PDS, the measured 
hard X--ray time lag also shows very similar characteristics within each 
transition
and between the transitions, but are markedly different between the transition
and the soft state. To illustrate this point, the results for Observations 
\#3, \#6, and \#15 (one in each phase) are summarized in Figure~8.
During the transition, hard X--rays clearly lag behind soft ones. The 
measured time lag shows a decreasing trend toward high frequency, 
confirming the Ginga results (\cite{miyamoto1988}). It show a peak at the 
QPO frequency --- most apparent between the soft and medium bands --- showing 
an additional time delay associated with the phenomenon. The time lag 
increases with 
photon energy. To quantify it, we defined an ``effective'' time lag for each 
energy band by simply averaging the results in the frequency range of 1--10 
Hz, where error bars are small. The results are plotted in Figure~9 for all 
observations. We also define an ``effective'' energy for each energy band,
\begin{equation}
E_{eff} = \frac{\int di \int E R(i,E) S(E) dE}{\int di \int R(i,E) S(E) dE},
\end{equation}
where S(E) is the photon flux at energy E, and R(i,E) is the detector response
matrix that distributes photons at energy E to counts in each pulse-height 
channel i; energy integrals are computed over a chosen energy band, while
pulse-height channel integrals are over all channels. Using the observed 
photon spectra for Phase 1 (\cite{cuietal1997a})
and PCA response matrices, the effective energy is computed for each energy 
band. Roughly, they are $\sim$3 keV, $\sim$9 keV, and $\sim$27 keV for the 
soft, medium, and hard bands, respectively, depending only weakly on the exact
spectral shape. Therefore, the time lag scales with photon energy roughly as 
$log(E_1/E_0)$ 
during the transitions, where $E_0$ and $E_1$ represent the effective energies
for any two energy bands. In the soft state, the time lags become hardly 
measurable between the same energy bands, and the logarithmic scaling with 
photon energy also breaks down.

\subsubsection{Coherence Function}

The coherence function is a measure of linear correlation between the two 
energy bands of interest (\cite{bendat1986}). In the noiseless case, it is 
defined as
\begin{equation}
\gamma(f) = \frac{<C(f)><C^*(f)>}{<PDS_1(f)><PDS_2(f)>},
\end{equation}
where $PDS_1(f) = F^*_1(f) F_1(f)$ and $PDS_2(f) = F^*_2(f) F_2(f)$ are the
PDSs for energy bands \#1 and \#2, respectively. Therefore, the coherence
function takes value in the range of 0 to 1, with 0 being no correlation
and 1 perfect linear correlation. In reality, noises
due to photon counting statistics are always present. A detailed treatment 
of such cases was presented and discussed thoroughly in Vaughan \& Nowak 
(1997). 

We calculated the coherence function, as well as its variance, by following 
the recipe presented in Vaughan \& Nowak (1997), which {\it only} applies to 
the cases of high signal power and high coherence. As shown in Figures 2, 3,
7, the signal power diminishes rapidly at high frequencies, so the coherence
function was calculated only at low frequencies. Figure~10 summarizes the 
results for Observations \#3, \#6, and \#15, as an example to show typical 
characteristics for each of the three phases. During the transition, the 
coherence function indicates a good linear correlation between the soft and 
medium bands, much less so between the soft and hard bands. Nearly perfect
linear correlations are observed between all three energy bands in the soft 
state.

As for the time lag, 
we averaged the coherence function between 1 and 10 Hz for all observations, 
and plotted the results in Figure~11. Clearly, the coherence function jumps 
around between observations during the transition, but remains quite stable 
in the soft state.

\section{DISCUSSION}

The observed PDS shows apparent evolution during the spectral transition.
In the hard state, the PDS of Cyg~X--1 can be characterized by a white noise 
component that extends up to $\sim$0.04--0.4 Hz where it breaks into roughly
a single power--law (review by \cite{vandk1995}; \cite{bellonietal1996}). 
During the hard--to--soft transition, the white noise weakens, and the cutoff 
frequency moves up to 1--3 Hz. In addition, a low--frequency 1/f component 
appears (see Fig~2, also \cite{bellonietal1996} and \cite{cuietal1997a}). As 
the source approaches the soft state, the 1/f noise strengthens until it 
dominates the PDS in the soft state (see Fig.~3). This evolution 
sequence completely reverses during the soft--to--hard transition (see Fig.~4). 
The absence of the 1/f noise in the hard state and its dominance in the soft
state seems to suggest that it is positively correlated with the {\it disk} 
mass accretion rate which is lower in the hard state and higher 
in the soft state (cf. \cite{zhang1997}). Therefore, it likely originates in 
the accretion disk. It can be produced by the fluctuation in the 
local accretion rate as a result of small random fluctuation in the viscosity 
(\cite{kazanas1996}), or by the superposition of random accretion 
``shots'' with long lifetimes (cf. \cite{belloni1990}).
The remarkable repeatibility of the properties of Cyg X-1 as observed in
recurrences of its various states and the similarity of the observed X-ray
properties during the transitions from the hard to soft and soft to hard
states implies an orderly dependence of the physical processes on some 
parameter. What are these processes and the possibilities for the parameter?

The hard X--ray emission from Cyg~X--1 in the hard state and in the 
transition state is probably the product of thermal
Comptonization. Although the exact geometry of the Comptonizing hot corona is 
still unknown --- many have been proposed --- there is evidence that it is 
present only in the vicinity of the black hole for X--ray binaries (e.g.,
\cite{gieretal1996}). The corona could be formed due to advection--dominated 
accretion flows (ADAF) (e.g., \cite{narayan1994}), or as a post--shock 
region in a centrifugally supported shock (e.g., \cite{chakrabarti1995}). It 
is still highly controversial if a shock can be formed, but the essential 
element is a Comptonizing region for both models. The ADAF model does provide 
an explanation of the different states (cf. \cite{narayan1996b}). According to
this model, the standard optically thick, geometrically thin disk 
(\cite{shakura1973}) is only present in 
the outer region, and is truncated near the black hole where an 
optically--thin hot ``corona'' is formed. The ``corona'' rotates at 
sub--Keplerian speed, and dissipates angular momentum via viscous processes, 
so behaves just like the thin disk. However, the gas density in the 
``corona'' is so low that radiation becomes an inefficient cooling mechanism.
Consequently, gravitational energy released in the accretion process heats up
the gas, and advects with it into the black hole. It was shown that when the 
mass accretion rate is below some critical value, the corona is large and the 
thin disk is far away from the black hole --- this is the low state. As the 
accretion rate increases, the inner edge of the disk moves inwards and the
corona shrinks, due to the increased local Compton cooling efficiency; 
the source is on its way to the high state. When the accretion rate exceeds 
the critical value, this process continues until the inner edge of the disk 
reaches the last stable orbit --- this is the high state. 

Is the ADAF model consistent with our results? The answer is yes, at least 
qualitatively. In previous work (\cite{cuietal1997a}), we speculated that 
the white noise 
perhaps originates near the black hole where dynamical time scale is short 
compared to the frequency range of the observed power. It is then 
filtered by the hot corona to produce the characteristic break ($f_{b2}$) on 
the PDS. The break frequency would then be determined by the characteristic 
photon escape time through the corona, i.e. by its physical size 
(\cite{cuietal1997a}; also a 
model in \cite{hua1996}). Therefore, the increase in the break frequency as 
the source approaches the soft state is consistent with the corona becoming 
smaller, which would also explain the softer energy spectrum in the soft 
state (\cite{cuietal1997a}). The temporal evolution of the corona 
during the transitions can then be illustrated by the change in the break 
frequency. 

Such a scenario is strongly supported by the measurement of hard X--ray time 
lag and coherence between different energy bands. The logarithmic scaling
of the observed time lag with photon energy are consistent with the predictions
of the thermal Comptonization in the corona (e.g., \cite{payne1980};
\cite{hua1996}; \cite{kazanas1996}). The much smaller lag times of the soft
state do not support scaling as the log of the photon energy, implying that
the large high temperature corona is not present. The loss of coherence 
during the transition (see Fig.~11) can also be accounted for by a varying 
corona. The coherence function being near unity in the soft state
rules out models invoking multiple, uncorrelated emission regions to account 
for X--ray variability on different time scales (cf. \cite{vn1997}). 

However, the current version of the ADAF model cannot satisfactorily describe
as luminous an X--ray source as Cyg~X--1 in the hard state. The maximum 
luminosity 
allowed by the model is $\sim 0.05\alpha^2 L_{E}$ (\cite{narayan1996a}), 
where $\alpha$ is the constant that describes viscosity in the standard 
thin--disk model (\cite{shakura1973}), and $L_E$ is the Eddington luminosity. 
For a Cyg~X--1 black hole of about $10M_{\odot}$, the measured luminosity is
about $0.03 L_E$ (\cite{zhang1997}), which would require $\alpha$ to be on 
the order of unity. Thus the reason for the hard state configuration is 
uncertain, but the spectral and temporal properties are qualitatively
in accord with several models, to the degree they are specified.

Alternatively, a shock could be formed close to the black hole in Cyg~X--1. 
Here, the post--shock region provides Comptonizing hot electrons, so there 
is essentially no difference between the two models for the hard state. 
In the soft state, however, Chakrabarti \& Titarchuk (1995) predicts the 
formation of a relativistic electron flow rushing towards the black hole, 
due to much more efficient Compton cooling. The bulk 
motion of the flow up--scatters soft photons to produce the observed hard 
X--ray emission.

With a varying corona, it is natural to ask if the detected QPO during
the transition is of coronal origin. As discussed in Section 2.1.1, there
is strong evidence that the observed QPO characteristics are related to hard
X--ray or coronal properties: the QPO amplitude increases as the energy 
spectrum becomes harder; both the QPO frequency and amplitude are correlated 
with the PDS break frequency ($f_{b2}$); and finally there is an extra time 
delay (of hard X--rays) at the QPO frequency. Therefore, the QPO is likely 
associated with some resonant oscillation in the corona. If the hot electron
corona is simply a post--shock region, it would not be 
hard to imagine the presence of shock--induced oscillations. 

The observed PDS during the transition is very similar to that of some 
soft X--ray transients, such as GX~339--4 and GS~1124--68, often observed in 
a so--called ``very high state'' (VHS; Miyamoto et al. 1991; Ebisawa et 
al. 1994; Belloni et al. 1997). The VHS is characterized by the 
presence of QPO in the frequency range similar to ours here. Often, both the 
low--frequency red noise and white noise are present. Moreover, the QPO is 
thought to be associated with the oscillations in the corona that are 
triggered by a radiation pressure feedback loop, assuming the X-ray luminosity
is super--Eddington in this state (\cite{vandk1995}). However, the QPO of 
Cyg~X-1 during the transition cannot be due to radiation pressure because the 
X-ray luminosity is only a few percent of the Eddington luminosity. Recently,
Belloni et al. (1997) showed that a QPO at 6.7 Hz was present in GS~1124-68 
during the transition from its high state to low state. This bears remarkable
similarity to what we see in Cyg~X-1.

\section{Conclusion}

Based on our results, we conclude that both the white noise and low--frequency
1/f noise are intrinsic to soft ``seed'' photons. These intrinsic temporal
properties are modified by the corona through Compton scattering process.
Therefore, the change in the high--frequency characteristics (e.g., $f_{b2}$) 
is interpreted as evidence for a ``fluctuating'' corona during the spectral 
transitions. This scenario is supported by the measurement of hard X--ray time
lags and coherence functions during the transition and in the soft state. 
The time lag scales roughly logarithmically with emerging photon energy, as 
predicted by thermal Comptonization models, thus is likely to be produced by 
such a process in the corona. This scaling relationship does not hold for the 
soft state, suggesting that new processes might be involved in the X--ray 
production.
 
The QPO persists during both the hard--to--soft and soft--to--hard 
transitions, but moves around between observations. For Cyg~X--1, it is 
likely a coronal phenomenon because its centroid frequency and strength
appear to be correlated with the hard X--ray properties, such as the energy 
spectrum, the PDS break frequency, and an extra time delay, that are tied to 
the corona. The presence of the QPO makes the PDS during the transition very 
similar to those of soft X--ray transients in the VHS. 

Our results are consistent with many of the qualitative features predicted by
the ADAF model. Recently, the results from the study of long--term spectral
evolution of Cyg~X--1, based on the ASM and BATSE data, provide further
evidence for the motion of the inner edge of the thin disk during the
spectral transitions (\cite{zhang1997}), which is an important prediction of 
the model. However, the current version of the model still requires extreme 
values of $\alpha$ parameter to account for the observed X--ray luminosity of
Cyg~X--1 (\cite{narayan1996a}). On the other hand, the bulk motion of 
relativistic electrons may indeed play a vital role in the soft state. 
The hard X-ray emission would then be produced by Comptonization due to the 
bulk motion of the electrons. Much can be learned by quantitatively applying 
these models to the spectral and temporal results from RXTE.

\acknowledgments
We wish to thank W.~Zhang for the dead--time correction code, X.~Hua and 
L.~Titarchuk for many useful discussions. This work is supported in part 
by NASA Contract NAS5--30612.

\clearpage

\begin{deluxetable}{lcc}
\footnotesize
\tablecolumns{3}
\tablewidth{0pc}
\tablecaption{{\it RXTE} Observations of Cyg~X--1}
\tablehead{
\colhead{Obs.} & \colhead{Obs. Time (UT)} & \colhead{PCA Live Time (s)}
}
\startdata
 1 & 5/22/96 17:44:00--19:48:00 & 4208 \nl
 2 & 5/23/96 14:13:00--18:07:00 & 7936 \nl
 3 & 5/30/96 07:46:00--08:44:00 & 2384 \nl
 4 & 6/04/96 20:21:00--21:42:00 & 3280 \nl
 5 & 6/16/96 00:00:00--00:40:00 &  816 \nl
 6 & 6/16/96 04:45:00--05:43:00 & 1312 \nl
 7 & 6/17/96 01:34:00--02:23:00 &  688 \nl
 8 & 6/17/96 04:46:00--05:43:00 & 1312 \nl
 9 & 6/17/96 07:58:00--09:07:00 & 2128 \nl
10 & 6/18/96 03:11:00--04:03:00 &  880 \nl
11 & 6/18/96 06:24:00--07:25:00 & 1680 \nl
12 & 6/18/96 09:36:00--10:45:00 & 2464 \nl
13 & 8/11/96 07:01:00--08:24:00 & 2688 \nl
14 & 8/11/96 15:08:00--15:51:00 & 1584 \nl
15 & 8/12/96 14:40:00--15:58:00 & 2114 \nl
\enddata
\end{deluxetable}

\clearpage

\begin{deluxetable}{lccccccccc}
\footnotesize
\tablecolumns{10}
\tablewidth{0pc}
\tablecaption{Characteristics of PDS during the Transitions}
\tablehead{
 & \multicolumn{5}{c}{Continuum} & \multicolumn{3}{c}{QPO} & \\
\cline{2-6} \cline{7-9} \\
\colhead{Obs.}&\colhead{$\alpha_1$}&\colhead{$f_{b1}$ (Hz)}&\colhead{$\alpha_2$}&\colhead{$f_{b2}$ (Hz)}&\colhead{$rms_c$ (\%)}&\colhead{$f_{qpo}$ (Hz)}&\colhead{FWHM (Hz)}&\colhead{$rms_{qpo}$ (\%)}&\colhead{$\chi_{\nu}/dof$}
}
\startdata
1a&$0.48^{+0.09}_{-0.07}$&$0.37^{+0.02}_{-0.08}$&$2.13^{+0.08}_{-0.06}$&$2.67^{+0.05}_{-0.04}$&$15.9^{+0.2}_{-0.2}$&$8.2^{+0.4}_{-0.5}$&$7.8^{+0.9}_{-0.9}$&$7.2^{+0.6}_{-0.7}$&$1.1/103$ \nl
1b&$0.64^{+0.07}_{-0.10}$&$0.28^{+0.04}_{-0.02}$&$2.05^{+0.12}_{-0.07}$&$2.66^{+0.05}_{-0.10}$&$14.3^{+0.2}_{-0.3}$&$8.8^{+0.2}_{-0.3}$&$6.8^{+1.4}_{-1.1}$&$5.9^{+0.9}_{-0.6}$&$1.2/103$ \nl
1c&$0.58^{+0.04}_{-0.04}$&$0.53^{+0.05}_{-0.04}$&$2.22^{+0.06}_{-0.04}$&$2.98^{+0.03}_{-0.03}$&$20.2^{+0.2}_{-0.2}$&$8.9^{+0.2}_{-0.3}$&$7.8^{+0.7}_{-0.6}$&$7.7^{+0.6}_{-0.5}$&$1.1/103$ \nl
2a&$0.28^{+0.08}_{-0.07}$&$0.23^{+0.04}_{-0.05}$&$1.93^{+0.19}_{-0.06}$&$0.96^{+0.09}_{-0.02}$&$16.4^{+0.3}_{-0.4}$&$3.6^{+0.3}_{-0.2}$&$6.9^{+0.3}_{-0.2}$&$16.9^{+0.6}_{-0.6}$&$1.6/103$ \nl
2b&$0.36^{+0.05}_{-0.06}$&$0.39^{+0.11}_{-0.05}$&$2.14^{+0.10}_{-0.08}$&$1.25^{+0.03}_{-0.02}$&$16.5^{+0.4}_{-0.3}$&$4.0^{+0.2}_{-0.2}$&$7.2^{+0.1}_{-0.2}$&$15.9^{+0.5}_{-0.6}$&$1.8/103$ \nl
2c&$0.49^{+0.18}_{-0.12}$&$0.11^{+0.03}_{-0.03}$&$2.08^{+0.08}_{-0.07}$&$1.07^{+0.03}_{-0.02}$&$17.1^{+0.4}_{-0.3}$&$3.8^{+0.2}_{-0.2}$&$7.1^{+0.1}_{-0.2}$&$16.0^{+0.5}_{-0.6}$&$1.2/103$ \nl
3&$0.68^{+0.07}_{-0.08}$&$0.32^{+0.09}_{-0.03}$&$1.97^{+0.05}_{-0.04}$&$2.81^{+0.03}_{-0.03}$&$15.0^{+0.1}_{-0.2}$&$9.4^{+0.2}_{-0.2}$&$6.6^{+0.8}_{-0.8}$&$5.8^{+0.4}_{-0.4}$&$2.1/103$ \nl
\tablevspace{1mm}
\cline{1-10}
\tablevspace{1mm}
13&$0.51^{+0.07}_{-0.04}$&$0.64^{+0.06}_{-0.11}$&$2.3^{+0.1}_{-0.1}$&$2.20^{+0.03}_{-0.04}$&$15.6^{+0.4}_{-0.4}$&$5.1^{+0.3}_{-0.2}$&$9.1^{+0.2}_{-0.2}$&$13.2^{+0.7}_{-0.7}$&$1.8/103$ \nl
14&$0.78^{+0.04}_{-0.04}$&$0.79^{+0.06}_{-0.04}$&$2.09^{+0.04}_{-0.03}$&$3.45^{+0.10}_{-0.04}$&$18.0^{+0.1}_{-0.1}$&$12.3^{+0.6}_{-0.8}$&$7^{+2}_{-2}$&$3.0^{+0.7}_{-0.5}$&$1.0/103$ \nl
15&$0.64^{+0.04}_{-0.05}$&$0.70^{+0.08}_{-0.04}$&$2.3^{+0.2}_{-0.2}$&$3.45^{+0.05}_{-0.05}$&$14.2^{+0.3}_{-0.3}$&$9.1^{+0.9}_{-0.9}$&$14^{+1}_{-2}$&$7^{+1}_{-1}$&$1.2/103$ \nl
\enddata
\end{deluxetable}

\clearpage

\begin{deluxetable}{lccccccccc}
\footnotesize
\tablecolumns{9}
\tablewidth{0pc}
\tablecaption{Characteristics of PDS in the Soft State}
\tablehead{
 & \multicolumn{4}{c}{Continuum} & \multicolumn{3}{c}{QPO} & \\
\cline{2-5} \cline{6-8}
\colhead{Obs.}&\colhead{$\alpha_1$}&\colhead{$\alpha_2$}&\colhead{$f_{b}$ (Hz)}&\colhead{$rms_c$ (\%)}&\colhead{$f_{qpo}$ (Hz)}&\colhead{FWHM (Hz)}&\colhead{$rms_{qpo}$ (\%)}&\colhead{$\chi_{\nu}/dof$}
}
\startdata
4&$0.88^{+0.01}_{-0.01}$&$2.35^{+0.06}_{-0.06}$&$13.3^{+0.4}_{-0.4}$&$18.4^{+0.1}_{-0.2}$&$6.2^{+0.3}_{-0.3}$&$5^{+1}_{-1}$&$4.8^{+0.7}_{-0.6}$&$1.1/104$ \nl
5&$1.01^{+0.04}_{-0.04}$&$2.4^{+0.4}_{-0.2}$&$13.5^{+0.7}_{-0.8}$&$18.6^{+0.4}_{-0.5}$&$3^{+2}_{-3}$&$13^{+3}_{-3}$&$9^{+3}_{-3}$&$0.7/104$ \nl
6&$1.01^{+0.02}_{-0.01}$&$2.4^{+0.1}_{-0.1}$&$14.0^{+0.7}_{-0.5}$&$18.7^{+0.2}_{-0.2}$&$6.5^{+0.7}_{-0.7}$&$6^{+2}_{-2}$&$4^{+1}_{-1}$&$1.4/104$ \nl
7&$0.95^{+0.01}_{-0.01}$&$2.6^{+0.2}_{-0.1}$&$12.8^{+0.6}_{-0.5}$&$16.6^{+0.2}_{-0.2}$&\nodata&\nodata&\nodata&$1.3/107$ \nl
8&$0.93^{+0.01}_{-0.01}$&$2.14^{+0.08}_{-0.06}$&$11.8^{+0.6}_{-0.4}$&$20.2^{+0.2}_{-0.2}$&$6.6^{+0.5}_{-0.3}$&$1.2^{+0.8}_{-0.7}$&$1.8^{+0.4}_{-0.4}$&$1.2/104$ \nl
9&$1.01^{+0.01}_{-0.01}$&$1.88^{+0.02}_{-0.02}$&$6.4^{+0.2}_{-0.3}$&$23.2^{+0.2}_{-0.2}$&\nodata&\nodata&\nodata&$3.0/107$ \nl
10&$0.97^{+0.03}_{-0.03}$&$2.3^{+0.2}_{-0.1}$&$12.8^{+0.6}_{-0.8}$&$17.2^{+0.2}_{-0.3}$&$7.7^{+0.7}_{-0.9}$&$10^{+3}_{-2}$&$6^{+2}_{-1}$&$0.9/104$ \nl
11&$0.99^{+0.01}_{-0.01}$&$2.6^{+0.1}_{-0.1}$&$13.6^{+0.5}_{-0.4}$&$17.5^{+0.2}_{-0.2}$&$6.5^{+0.4}_{-0.4}$&$1.6^{+1.5}_{-0.9}$&$1.6^{+0.5}_{-0.4}$&$1.0/104$ \nl
12&$0.98^{+0.01}_{-0.01}$&$1.94^{+0.04}_{-0.03}$&$8.8^{+0.3}_{-0.2}$&$22.1^{+0.2}_{-0.2}$&\nodata&\nodata&\nodata&$1.5/107$ \nl
\enddata
\end{deluxetable}

\clearpage
\begin{figure}[t]
\epsfxsize=350pt \epsfbox{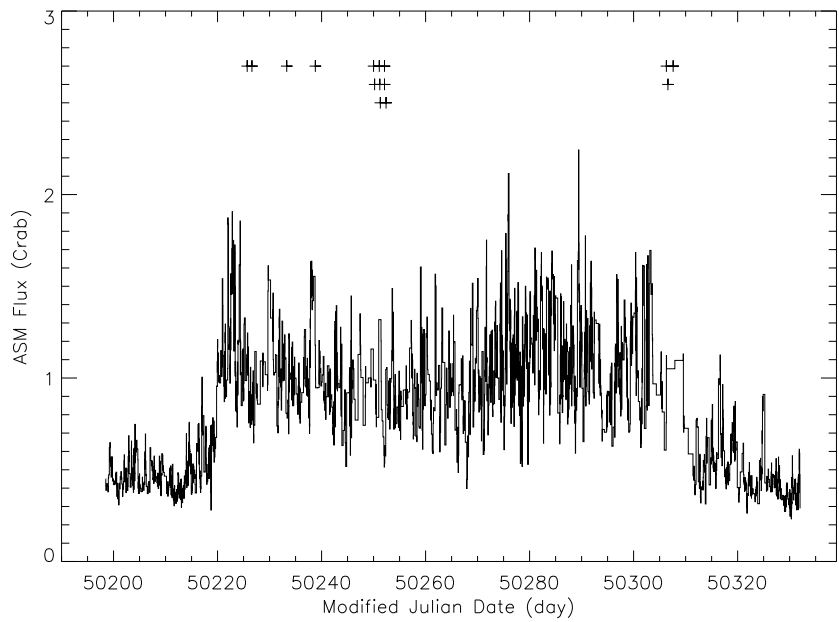}
\caption{ASM light curve of Cyg~X--1. It comprises measurements from 
individual ``dwells'' with 90--second exposure time. The crosses indicate 
when the {\it RXTE} observations were made. MJD 50213.0 corresponds to 
1996 May 10 0 h UT.}
\end{figure}

\clearpage
\begin{figure}[t]
\epsfxsize=350pt \epsfbox{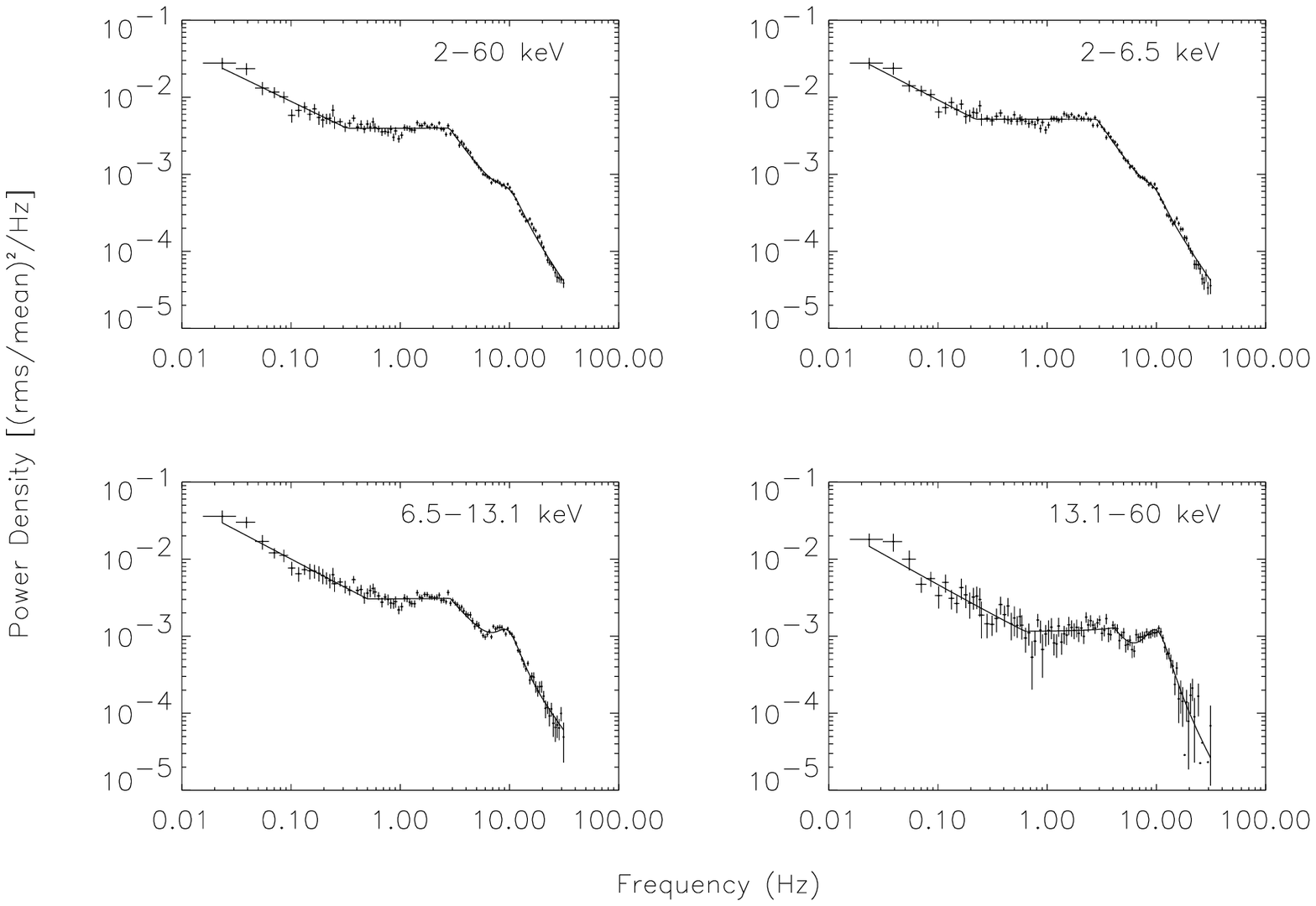}
\caption{Power density spectrum of Observation \#3 (see Table~1) in the
energy bands indicated. It is a representive of the hard--to--soft transition.
The data are logarithmically rebinned to reduce scatter at high frequencies.
The solid line shows the best--fit model.}
\end{figure}

\clearpage
\begin{figure}[t]
\epsfxsize=350pt \epsfbox{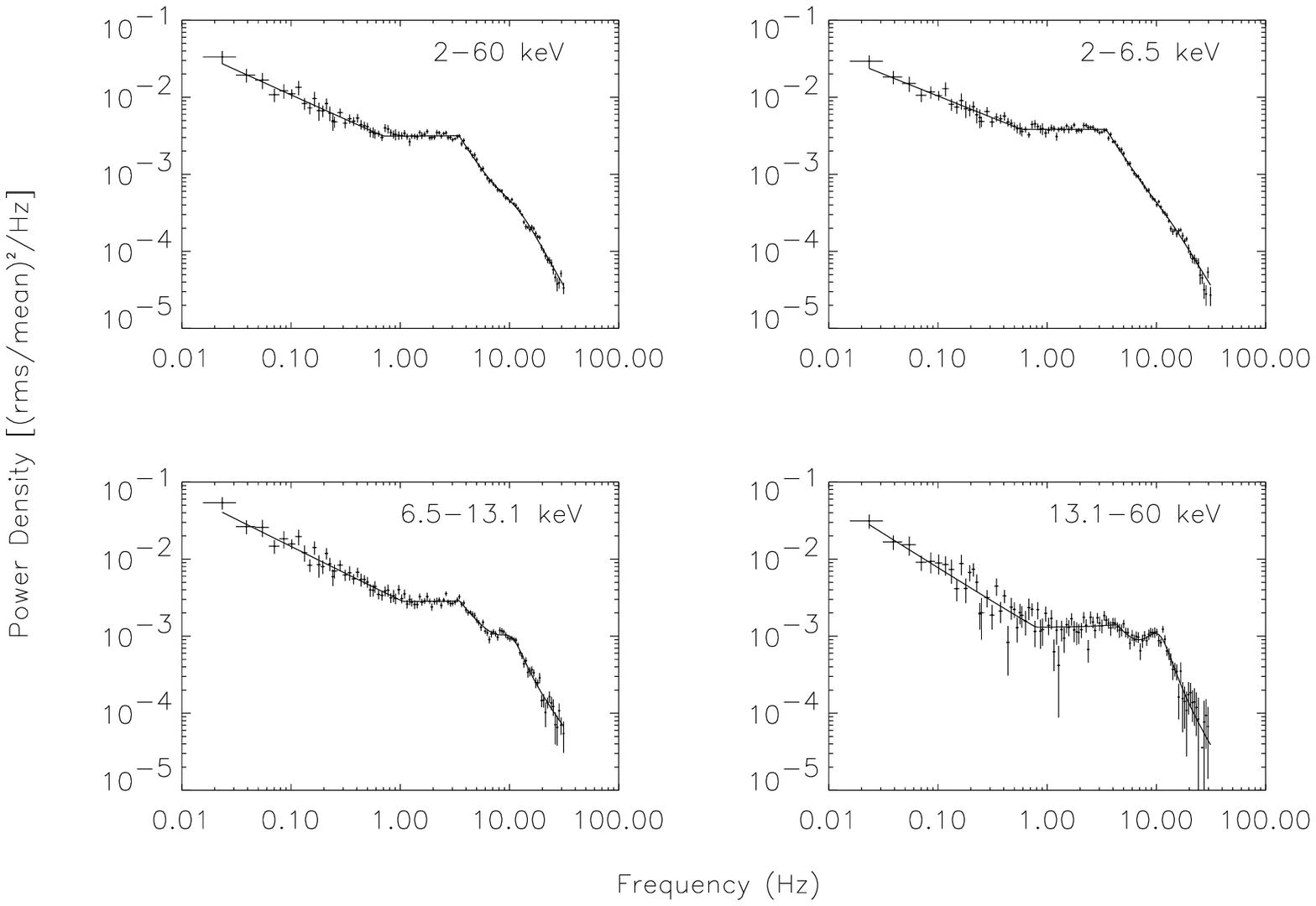}
\caption{Power density spectrum of Observation \#15 (see Table~1) in the
energy bands indicated. It is a representive of the soft--to--hard transition.
The data are logarithmically rebinned to reduce scatter at high frequencies.
The solid line shows the best--fit model.}
\end{figure}

\clearpage
\begin{figure}[t]
\epsfxsize=350pt \epsfbox{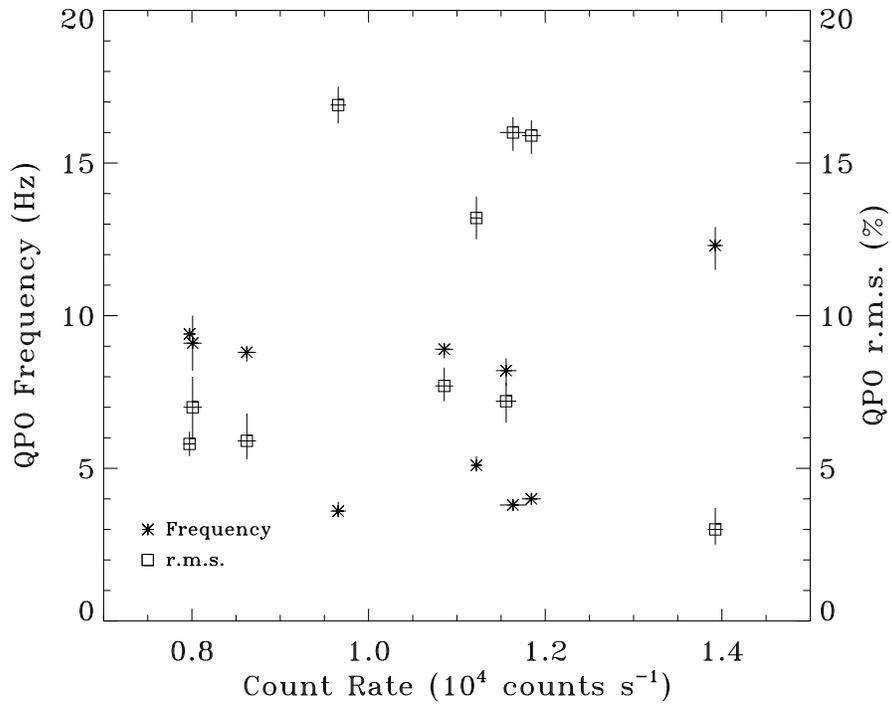}
\caption{QPO characteristics at various source count rates. No apparent
correlation can be seen.}
\end{figure}

\clearpage
\begin{figure}[t]
\epsfxsize=350pt \epsfbox{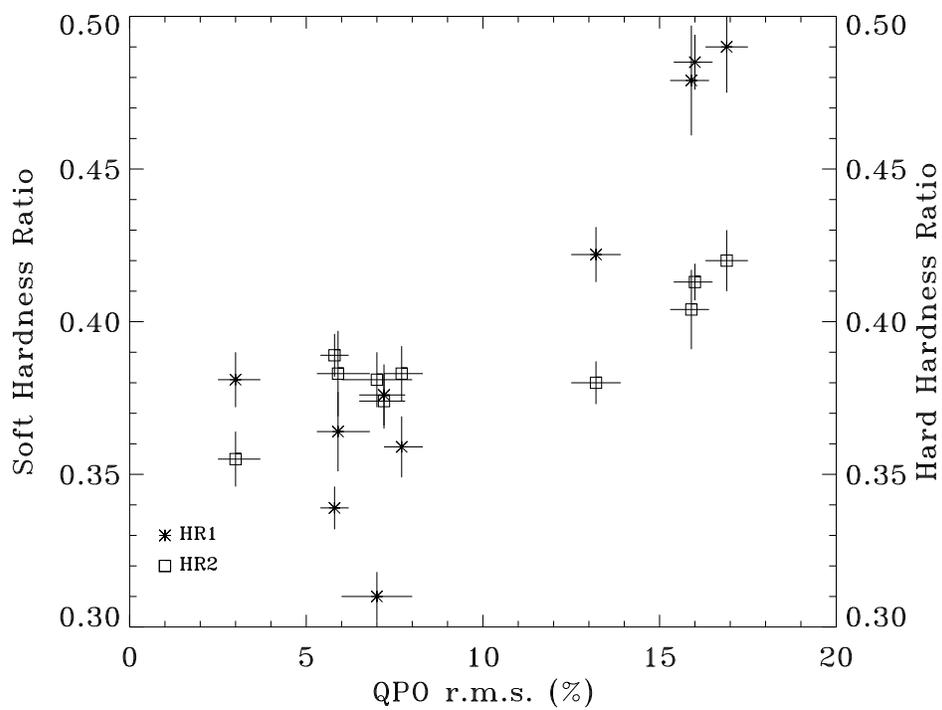}
\caption{Spectral dependence of the QPO amplitude. HR1 and HR2 are defined
in Section 2.1.1.}
\end{figure}

\clearpage
\begin{figure}[t]
\epsfxsize=350pt \epsfbox{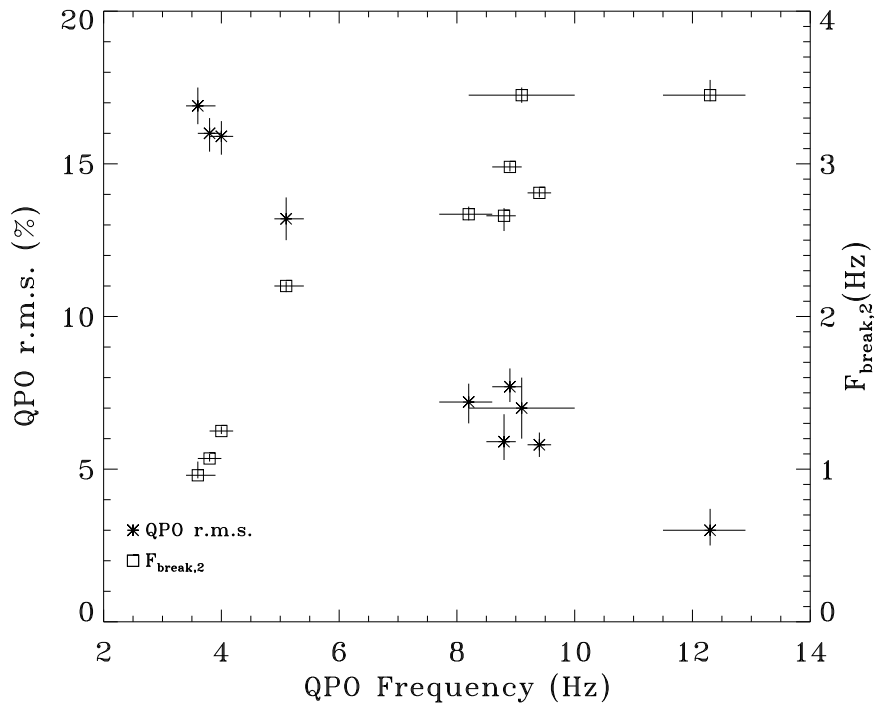}
\caption{QPO frequency, QPO amplitude, the PDS break frequency, and 
correlations between them.}
\end{figure}

\clearpage
\begin{figure}[t]
\epsfxsize=350pt \epsfbox{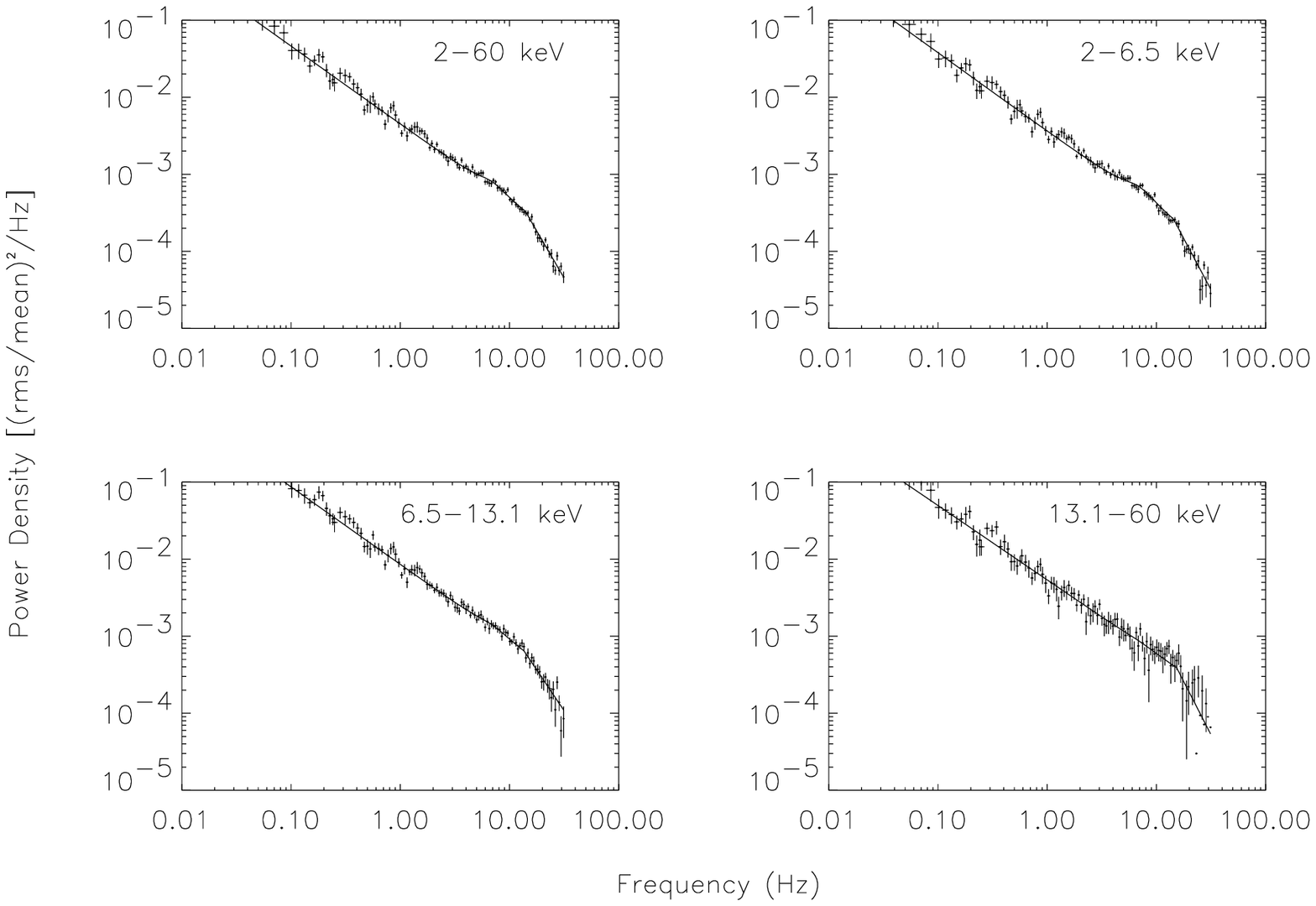}
\caption{Power density spectrum of Observation \#6 (see Table~1) in the
energy bands indicated. It is a representive of the soft state. The data 
are logarithmically rebinned to reduce scatter at high frequencies.
The solid line shows the best--fit model.}
\end{figure}

\clearpage
\begin{figure}[t]
\epsfxsize=350pt \epsfbox{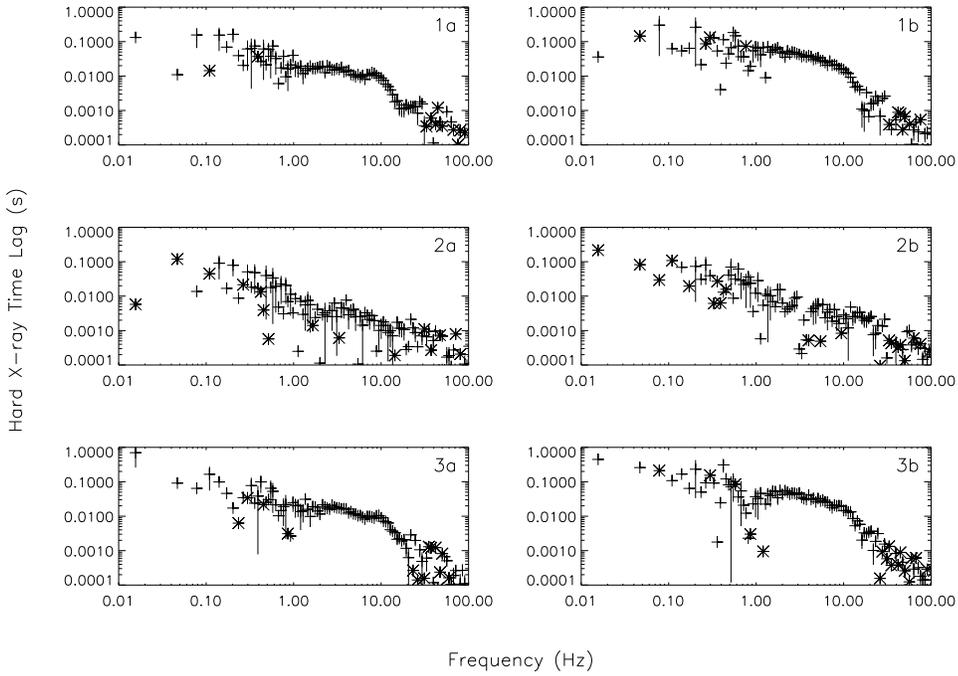}
\caption{Hard X--ray time lag. Three panels on the left (top to bottom) show 
in crosses the measurements between the soft and medium bands for Observations
\#3, \#6, and \#15, respectively, and the right panels between the soft and 
hard bands. The error bars that extend to negative values are not plotted. The
asterisks indicate negative values. }
\end{figure}

\clearpage
\begin{figure}[t]
\epsfxsize=350pt \epsfbox{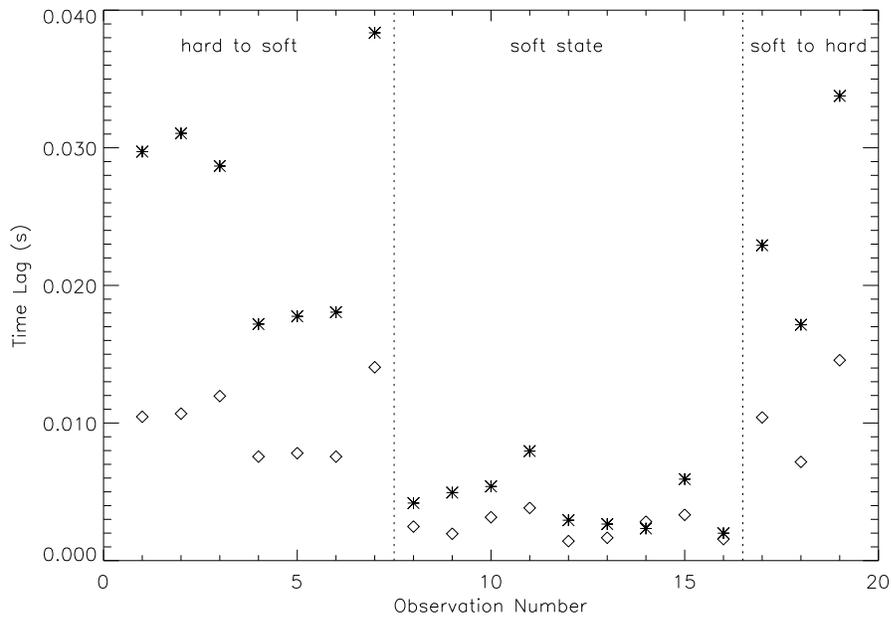}
\caption{``Effective'' time lags for all observations (see definition in
Section 2.2.1). The diamonds show the measurements between the soft
and medium bands, and the asterisks between the soft and hard bands. Three 
distinct phases during the entire period are marked, and separated by 
dotted--lines.}
\end{figure}

\clearpage
\begin{figure}[t]
\epsfxsize=350pt \epsfbox{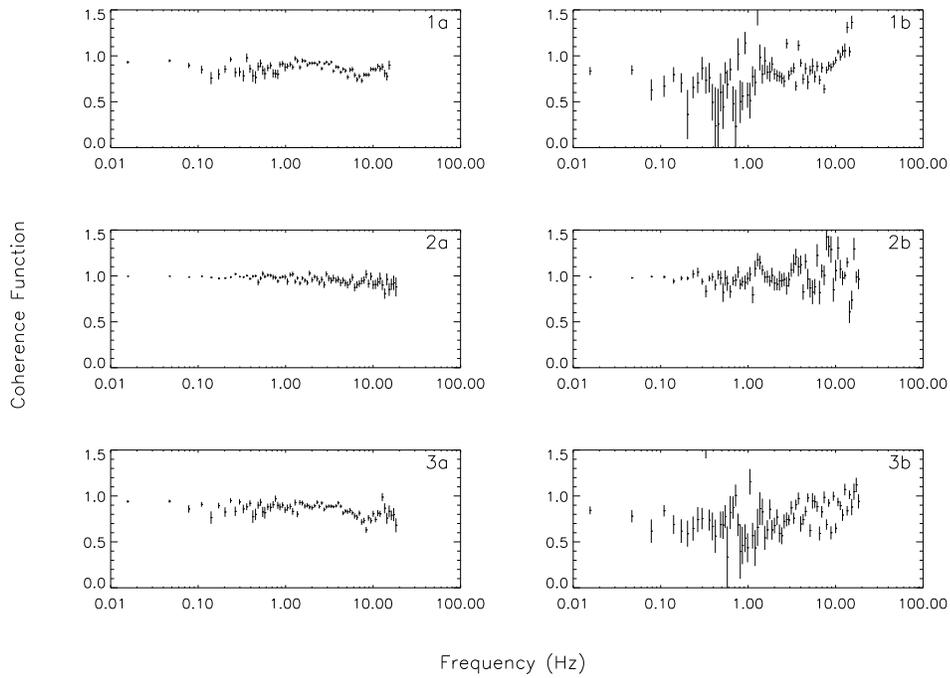}
\caption{Coherence function. Three panels on the left (top to bottom)
show the measurements between the soft and medium bands for Observations \#3,
\#6, and \#15, respectively, and the right panels between the soft
and hard bands.}
\end{figure}

\clearpage
\begin{figure}[t]
\epsfxsize=350pt \epsfbox{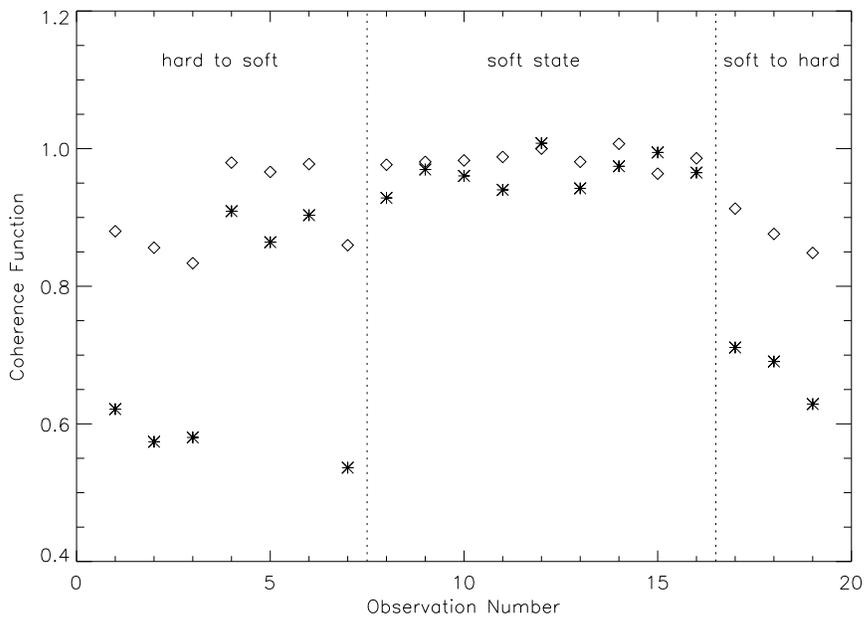}
\caption{Average coherence function for all observations (see text). The 
diamonds show the measurements between the soft and medium bands, 
and the asterisks between the soft and hard bands. Three 
distinct phases during the entire period are marked, and separated by 
dotted--lines.}
\end{figure}

\end{document}